\documentclass[fleqn,10pt]{wlscirep}

\title{Adaptation to criticality through organizational invariance in embodied agents}

\author[1,2,3,*]{Miguel Aguilera}
\author[1,4]{Manuel G. Bedia}
\affil[1]{Dept. of Computer Science, University of Zaragoza, Zaragoza, Spain}
\affil[2]{IAS-Research Center for Life, Mind, and Society, University of the Basque Country, Donostia-San Sebastián, Spain}
\affil[3]{Institute for Cross-Disciplinary Physics and Complex Systems, Palma, Spain}
\affil[4]{Arag\'on Institute of Engineering Research, Zaragoza, Spain}
\affil[*]{sci@maguilera.net}

\begin{abstract}
Many biological and cognitive systems do not operate deep within one or other regime of activity. Instead, they are poised at critical points located at phase transitions in their parameter space. The pervasiveness of criticality suggests that there may be general principles inducing this behaviour, yet there is no well-founded theory for understanding how criticality is generated at a wide span of levels and contexts.
In order to explore how criticality might emerge from general adaptive mechanisms, we propose a simple learning rule that maintains an internal organizational structure from a specific family of systems at criticality.
We implement the mechanism in artificial embodied agents controlled by a neural network maintaining a correlation structure randomly sampled from an Ising model at critical temperature.
Agents are evaluated in two classical reinforcement learning scenarios: the Mountain Car and the Acrobot double pendulum. In both cases the neural controller appears to reach a point of criticality, which coincides with a transition point between two regimes of the agent's behaviour.
These results suggest that adaptation to criticality could be used as a general adaptive mechanism in some circumstances, providing an alternative explanation for the  pervasive presence of criticality in biological and cognitive systems.
\end{abstract}
\begin{document}

\flushbottom
\maketitle
\thispagestyle{empty}

\section*{Introduction}

Generally, the behaviour of biological and cognitive systems is not steadily poised at one phase or another. Instead, living beings operate around points of critical activity, at the boundary separating ordered and disordered dynamics. Critical activity, or \emph{criticality}, refers to a distinctive set of properties found at this transition. Some of these properties include the presence of a wide range of scales of activity and maximal sensitivity to external fluctuations \cite{salinas_phase_2001,salinas_scaling_2001}, facilitating systems at criticality to present optimal responses when facing complex heterogeneous environments \cite{hidalgo_information-based_2014}.
The surprising fact is that, unlike unanimated matter where critical transitions from order to disorder take place by fine-tuning of the parameters of the system \cite{salinas_phase_2001}, living systems appear to be ubiquitously poised near critical points \cite{mora_are_2011}. For instance, signatures of criticality have been detected in neural cultures \cite{schneidman_weak_2006}, immune receptor proteins \cite{mora_maximum_2010}, the network of genes controlling morphogenesis in fly embryos \cite{krotov_morphogenesis_2014} and bacterial clustering \cite{chen_scale-invariant_2012}. Indicators of critical behaviour have also been observed in the brain \cite{chialvo_critical_2014} and cognitive behavioural patterns \cite{van_orden_blue-collar_2012}.

Although these results suggest that general theoretical principles might underlie biological self-organization, there is no well-founded theory yet for understanding how living systems operate near critical points in a broad range of contexts. This compels us to ask what type of mechanisms are driving biological systems at a dauntingly diverse span of levels of organization to operate near critical points of activity. During the last couple of decades, the issue has been popularized as the `adaptation to the edge of chaos' \cite{kauffman_origins_1993} and different solutions have been tested through modelling approaches. However, as we have stated, this question is still unresolved and a general mathematical framework for understanding how living systems are driven to criticality is yet lacking.

If we had to synthesize the approaches that have been proposed so far to explain the presence of criticality spreading over such a wide range of natural systems, we could  broadly consider two general positions. On the one hand, we find approaches assuming that biological systems are self-tuned, either by learning or evolution, to regions of the parameter space displaying optimal fitness, and that these optimal points are often placed near critical points due to the functional advantages of critical behaviour\cite{beggs_neuronal_2003, hidalgo_information-based_2014}. In other words, criticality is understood as a by-product derived from the adaptive or survival processes of living systems. On the other hand, we find different views focused on the idea of self-organized criticality systems (SOC), in which criticality emerges spontaneously from simple local interactions, without fine tuning of the parameters of the model. Typically, SOC models exploit clever local rules (e.g. in cellular automata) that produce critical behaviour in a specific context \cite{bak_self-organized_1987, bak_punctuated_1993, jensen_self-organized_1998}.

Nevertheless both approaches present some weaknesses. Understanding criticality in biological systems as an indirect consequence of adaptation to external circumstances is not very explanatory and does not provide ways to test alternatives. In general, explanations that simply assume that they are the consequence of the `survival of the fittest' are often unfalsifiable. With regard to the second approach, it is well known that many SOC models are highly idealized and, in many cases, they are not able to capture the basic interactions of living systems, often failing to provide general explanations of the ubiquitous emergence of criticality\cite{watkins_25_2016}. For that reason, although both adaptive self-tuning and SOC models present interesting insights, they are generally applicable in a relatively narrow set of contexts or under highly idealized conditions.

In this paper, we explore an alternative approach to examine how a system can display critical activity in a wide variety of situations. We propose a model that, using only general local mechanisms, is aimed to adaptively maintain the behaviour of the system around a critical point of its parameter space by maintaining certain relational invariants. In other words, instead of thinking about criticality as a by-product of adaptation to complex environments or a spontaneous property of certain systems, we inquire into the possibility that biological systems might are equipped with adaptive mechanisms aimed to preserve an internal equilibrium near critical points while they interact with their environment. Among other outcomes, the existence of cheap learning mechanisms maintaining the parameters of a system around regions of criticality could drastically reduce the cost of searching large parametric spaces for finding fit solutions, or even generate interesting solutions in an unsupervised manner.

The paper is structured as follows.
First, we propose a novel method that appears to be capable of driving a system near critical points by maintaining certain relational invariants. In our case, these invariants are extracted from the correlation structure of well-known model operating at a critical point, where correlations scale with distance according to a power law function.
Second, we propose a simple learning rule maintaining this correlation invariance, and hypothesize that it could be used for driving systems in different contexts to operate near critical points.
We test the model using two classical examples of learning and control: the Mountain Car and a double pendulum. We show evidence suggesting that the general rule proposed here is able to drive adaptive agents with no free parameters towards critical points of operation. 
At the same time, the agents themselves are poised at points of behavioural transition, where they are able to exploit a broad span of dynamic possibilities available in their environment, suggesting a link between an internal search of critical points and the exploration of external behavioural points of transition.
Finally, we suggest further tests of criticality and discuss the limitations and possible generalization of this synthetic approach as a contribution towards understanding deeper principles governing biological and cognitive systems.

\section*{Model}

Inspired by the ideas described above, we present a novel simple mechanism designed to test whether a general adaptive system is able to drive a neural controller near criticality by imposing certain patterns in the organizational structure of the system. This focus on the system's organization instead of the mechanistic properties of its components is supported by the existence of well-known universality classes that provide a unified expression for families of systems operating under criticality \cite{kadanoff_more_2009}.

In physics, the concept of universality allows to group a great variety of different critical phenomena into a small number of universality classes in such a way that all systems belonging to a given universality class are essentially identical near the critical point. Thus, systems belonging to the same universality class, even if defined by very different material parameters or physical properties, have the same critical exponents characterizing diverging observables. For example, in different spin and percolation models, we find that the family of all bidimensional lattices (square, triangular, hexagonal and so forth) spatial correlations follow the asymptotic form $c(r) \propto 1/r^{\eta}$ near the phase transition point, where $\eta$ is the same for all lattice structures of dimension 2 in a particular model.

This surprising property provides a perspective on criticality in terms of universal relations, suggesting that we could model criticality using simple and non-specific mechanisms independently of the individual parameters of the system. Our hypothesis is the following: if all systems belonging to the same universality class present the same distribution of correlations at criticality, adjusting an arbitrary system to reproduce the same distribution of correlations might drive the system to a similar critical point. If, as we have said, in the neighborhood of critical points, critical exponents assume the same universal values for a particular class, it could be enough to use in our analysis a very simple (but nontrivial) model. Looking for generality, we use the least structured statistical model (i.e. a maximum entropy model) of a network of interacting units, constrained only by pairwise correlations between them. This is known as the Ising model in physics or the Boltzmann Machine in computer science \cite{ackley_learning_1985}. The interest of using it is that it is also one of the simplest models of criticality that that can be solved analytically\cite{onsager_crystal_1944}.

An Ising model can be specified as a neural network of $N$ binary variables only constrained by pairwise correlations. Units can have a value of $+1$ or $-1$ and are affected by local bias  $h_i$ and couplings  $J_{ij}$ between pairs of units. These parameters take continuous values and we assume couplings to be symmetric with $J_{ii}=0$. In order to simulate the behaviour of the model, units are updated sequentially in a random order using Glauber dynamics, by which each unit is activated with a probability that follows a sigmoid function:
\begin{equation}
 P(s_i(t+1)) = \Big[ 1 + e^{- \beta 2 H_i(t)  s_i(t+1) }\Big]^{-1}
 \label{eq:Glauber}
\end{equation}
where $H_i(s) = h_i + \sum_j J_{ij} s_j(t)$ is the effective field received by neuron $i$ summing the internal field and inputs from other neurons (and $2 H_i(s) s_i(t+1)$ is the energy difference required to flip the sign of unit $i$). The state of the model will be updated by sequentially applying Glauber dynamics (i.e. Equation~\ref{eq:Glauber}) each simulation step to all units of the network in a random order. When updated sequentially, an Ising model with symmetric couplings will reach an equilibrium maximum entropy distribution:
\begin{equation}
 P(s) =   \frac{1}{Z} exp \Bigg[ {\beta (\sum_i h_i s_i + \sum_{i < j} J_{ij} s_i s_j})\Bigg]
 \label{eq:Ising}
\end{equation}
where the distribution follows an exponential family $P(s) = \frac{1}{Z} e^{-\beta E(s)}$ and $Z$ is a normalization value. The energy $E(s)$ of each state of size $N$ is defined in terms of the bias  $h_i$  and couplings  $J_{ij}$  between pairs of units, with  $\beta = 1/({T k_B})$, being $k_B$ the Boltzmann's constant and $T$ the temperature of the system. Without loss of generality we can set an operating working temperature such that $\beta= 1$.

Following the intuition introduced above about universality classes, we are looking for a model that preserves certain structure in the correlations of the system. There is some experimental evidence showing that, given an Ising model near a critical point, one could build a family of models by learning correlations drawn at random from the original system, which will be poised near a critical point \cite{tkacik_spin_2009}. Inspired by this idea, we propose to reproduce and support criticality by maintaining a distribution of correlations of a particular universality class. Interestingly, the Ising model is a well-studied example of a universality class. In the case of bidimensional lattices, pairwise correlations follow the asymptotic form $c(r) \propto 1/r^{\eta}$, where $\eta=1/4$ \cite{salinas_scaling_2001}, and $r$ is the distance between units.

However, instead of restricting a model to a particular set of mechanisms or a given topology, we decide to design a learning rule that preserves a distribution of correlations of a critical point belonging to a specific universality class. This could capture some of the properties of that universality class without choosing a specific topology or parametrical configuration. The next goal would be to test whether this adaptive mechanism has the capability of driving an arbitrary system to a critical point. And, in case it were so, the system poised near a critical point should display interesting features of adaptive behaviour as maximal sensitivity or a wide dynamic range of behaviours.

In order to test this idea we design a  simple learning rule to adjust the parameters of an arbitrary Ising model to the desired distribution of correlations.
In a nutshell, the learning rule will operate as follows: 1) The distribution of correlations of a finite square lattice Ising model is calculated, 2) the new model is defined by assigning each neuron reference correlation values for its synapses randomly sampled from the previous distribution, 3) during learning, each neuron sorts its synapses by their correlation strength, and adjusts these correlations to the reference values assigned using an inverse Ising learning rule. We proceed now to explain these points in detail.

\begin{figure}[ht]
\begin{center}
 \includegraphics[width=7cm]{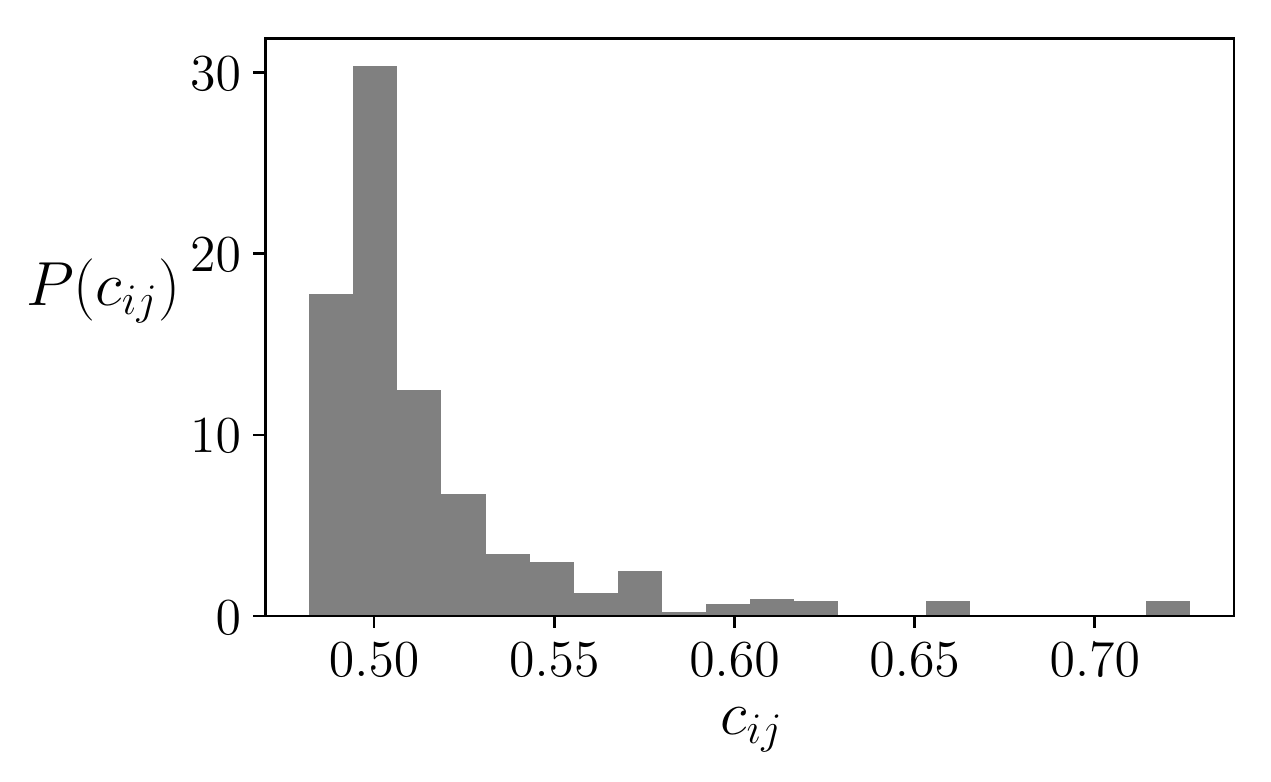}
\end{center}
\caption{Distribution of correlation values used for learning, generated from a 20x20 lattice Ising model at critical temperature.} 
\label{fig:Correlations}
\end{figure}

First, since the size of the models employed here is far from the thermodynamic limit, instead of directly using the diverging asymptotic form $c(r) \propto 1/r^{\eta}$, we approximate it by computing the correlation structure of a finite model operating at a known critical temperature. 
One of the few cases where the Ising model presents an exact solution is a model with zero fields and a bidimensional square lattice connectivity, in which a critical point appears at \mbox{$J= log{(1+\sqrt{2})}/(2 \beta)$} \cite{onsager_crystal_1944}. 
Exploiting this, we build a 20x20 square lattice Ising model operating at critical temperature with periodic boundary conditions to generate reference correlations to be used by the learning rule. 
We simulate the model using Glauber Dynamics, generating $10^6$ samples, after an initial run of $10^5$ updates starting from a random state. From this simulation, we obtain the distribution of correlations in the system $P(c_{ij})$, where $c_{ij}=\langle s_i s_j\rangle$, observed in Figure \ref{fig:Correlations}. Since the fields $h_i$ of all units are zero, the means $m_i = \langle s_i \rangle$ of all units are also zero.

Second, once the distribution of correlations has been obtained, we generate new models by adjusting their correlations to match the distribution $P(c_{ij})$. For doing so applying only local information, we assign each neuron a set of correlation values $c^*_{ik},k=1...N$ drawn at random from $P(c_{ij})$. At each step, we will compute the actual correlations between a neuron $i$ and its neighbours $j$ as $c^m_{ij}$, and generate reference values for learning by sorting $c^*_{ik}$ to match the order of $c^m_{ij}$. We will denote the sorted reference values as $c^*_{ij}$. All the values of  $m^*_{i}$ will be set equal to zero. The reason for sorting the values of the correlations is to give more flexibility to the rule, since we are only interested in maintaining a $c(r) \propto 1/r^{\eta}$ relation, independently of which connection holds each value.

For the third step, the problem is that it is not trivial finding which combination of $h_i$ and $J_{ij}$ generates a specific combination of $m_j$ and $c_{ij}$. This is known as the `inverse Ising problem', which can be solved by using a simple gradient descent rule \cite{ackley_learning_1985}:

\begin{equation}
 \begin{split}
  h_{i} \leftarrow h_{i} + \mu  (m^*_i - m_i^m )\\
  J_{ji} \leftarrow J_{ji} + \mu  ( c^*_{ij} - c_{ij}^m )     
 \end{split}
  \label{eq:learning}
\end{equation}
where $\mu$ is a constant learning rate, $m^*_{i}$ and $c^*_{ij}$ are the reference mean and correlations of the learning algorithm, and $m_{i}^m$ and $c_{ij}^m$ are the mean and correlations of the model for the current values of $h_i$ and $J_{ij}$. 
Generally, performing each learning step is computationally expensive, since it requires summing over all possible states of $s$, although approximate methods such as Monte Carlo sampling are generally used to speed up learning. Similarly, we compute the approximate correlations by simulating our networks using the Glauber dynamics in Equation~\ref{eq:Glauber} for a number of steps.

Now, before jumping to the main results, we will present a test of the model under idealized conditions and describe the embodied agents we use for evaluating the performance of the presented learning rule.

\subsection*{Testing adaptation to criticality in isolated networks}

As a demonstration of our method, we apply the learning rule to 10 different networks for sizes $N=4, 8, 16, 32, 64$ assigning them means and correlations drawn at random from the distribution found for the 20x20 lattice Ising model. For each network, we apply Equation~\ref{eq:learning} for learning $m^*_{i}$ and $c^*_{ij}$, estimating the actual $m_{i}^m$ and $c_{ij}^m$ with Glauber dynamics.

In order to simplify the process, we have made some operative decisions. For instance, since precision of learning is not important (the objective is to capture the overall distribution), we do not wait for convergence of the algorithm and simply update the learning rule $1000$ times. We use a learning rate $\mu=0.01$ and compute $1000 N$ samples for each learning step, being $N$ the size of the system. For simplicity also, instead of the critical temperature of the lattice Ising model, we set an arbitrary inverse temperature of $\beta=1$. Note that the choice of operating temperature is irrelevant, since it only implies a rescaling of the parameters that the algorithm will compensate.

For each network, we test if it is near a critical point by computing its heat capacity. A divergence in the heat capacity is a sufficient indicator of a continuous phase transition indicating the presence of a critical point with maximal sensitivity to parametrical changes \cite{salinas_phase_2001}.
In our model, the heat capacity is represented by $C(\beta) = - \beta \frac{\partial H}{\partial \beta}= \beta^2 ( \langle E^2(s) \rangle - \langle E(s) \rangle ^2)$, where $E(s) = - \sum_i h_i s_i - \sum_{i < j} J_{ij} s_i s_j$ is the energy of the Ising model and $H=-\sum_s P(s)\log{P(s)}$ is the entropy of the system. So we test if a system is at criticality by using the entropy $H$ of the system as an order parameter and looking for continuous phase transitions associated with critical points. We detect a continuous phase transition if the entropy presents a sharp but continuous transition in which the derivative of the entropy (the heat capacity) diverges as the system size increases.

\begin{figure}[ht]
\begin{center}
 \includegraphics{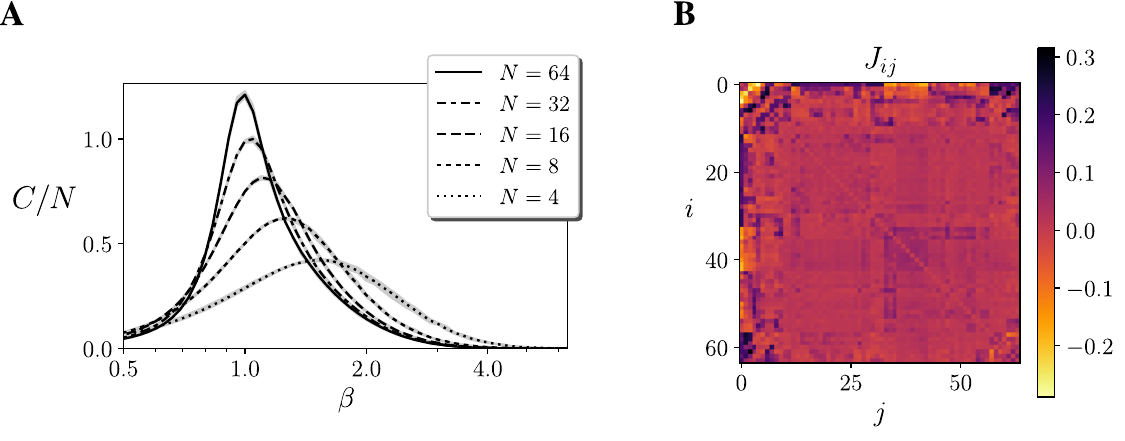}
\end{center}
\caption{\textbf{(A)} Divergence of the heat capacity in 10 models after learning random correlations sampled from Figure \ref{fig:Correlations}. Maximum and minimum values are shown by the grey area. \textbf{(B)} Distribution of the values of the coupling matrix $J_{ij}$ of a 64 units Ising model after learning correlations sampled from Figure \ref{fig:Correlations}. (the order of the nodes in the coupling matrix has been arranged using hierarchical clustering). Values of $J_{ij}$ correspond to the values in the colour bar.}
\label{fig:Test-Ising}
\end{figure}

We simulate each network for $10^5$ steps for different values of $\beta$, and we find that all the heat capacity of the 10 networks diverges at the operating temperature $\beta=1$ (Figure \ref{fig:Test-Ising}.A), showing values similar to those of the original lattice Ising model with periodic boundaries. Let us point out that although the distribution of correlations is similar to the lattice Ising model, the structure of the network is radically changed. Instead of the original ordered structure of a uniform lattice, we now have a disordered distribution of couplings $J_{ij}$ (Figure \ref{fig:Test-Ising}.B), including both positive and negative values. Also, each execution of the learning algorithm yields a completely different arrangement of values of couplings $J_{ij}$.

In the following section, we test the capacity of this learning rule for driving the neural controller of an embodied agent towards a critical point. In order to do so, we need to take into account the environment during learning. If we consider two interconnected Ising models (one being the neural controller and other being the environment) Equation~\ref{eq:learning} holds perfectly if we only apply it to the values of $i$ and $j$ corresponding to units of the neural controller. In our case, we do not use an Ising model as an environment but instead we use two classical examples in reinforcement learning with the goal of testing a more realist scenario. Therefore, our learning rule will be valid as long as the statistics of the environment can be approximated by an Ising model with an arbitrary number of units. Luckily, Ising models in the form of Boltzmann machines are universal approximators \cite{montufar_universal_2014} and the stationary distribution of any arbitrary environment can be approximated by an equivalent Ising model.

\begin{figure}
\begin{center}
 \includegraphics{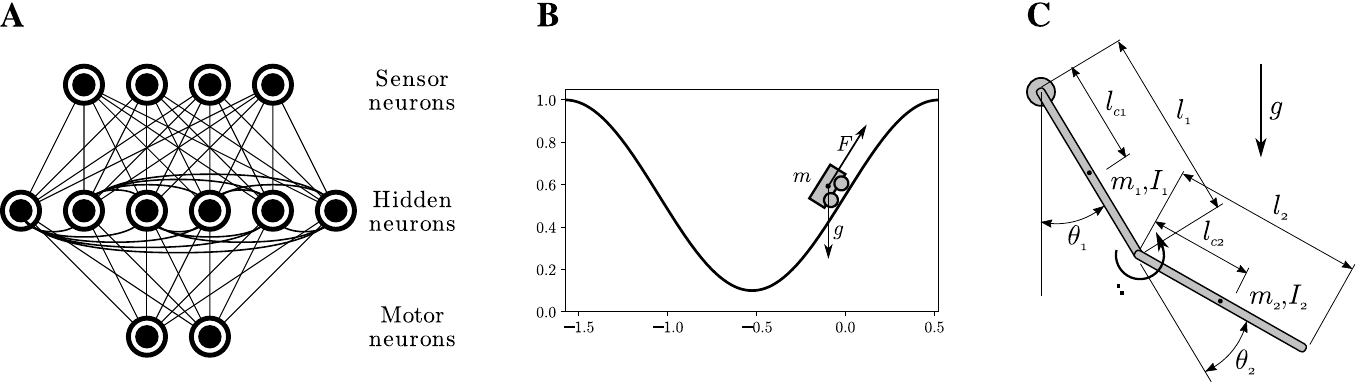}
\end{center}
\caption{(\textbf{A}) Structure of the embodied neural controller for a model with $N=12$ units and $N_h=6$ hidden units. (\textbf{B}) Mountain Car environment: an under-powered car that must drive up a steep hill by balancing itself to gain momentum. (\textbf{C}) Acrobot environment: an agent has to balance a double pendulum to reach the high part of the environment.} 
\label{fig:Embodiment}
\end{figure}

\subsection*{Embodied model: Mountain Car and Acrobot}
In order to evaluate the behaviour of the proposed learning rule, we test it in two embodied situations using the OpenAI Gym toolkit \cite{brockman_openai_2016}. 
We define a neural network consisting of an Ising model defined as in Equation~\ref{eq:Ising}, describing a network of $N=6 + H_h$ units, with $N_h$ hidden units, $2$ motor units and $4$ sensor units.
Motor units define the actions performed by the agents. In sensor units, the magnetic field of the unit is not a fixed parameter but it is be updated with the value of an external input $h_i=I_i$.
Sensor units and motor units are only connected to hidden neurons, while hidden neurons are connected to all other neurons (Figure \ref{fig:Embodiment}.A). We choose this configuration because it is widely used in neural networks and allowed recurrence in the connectivity of hidden units, although the architecture choice is not restrictive.
All units are assigned an reference $c^*_{ik}$ values (selected at random from distribution $P(c_{ij})$ shown in Figure \ref{fig:Correlations}), and all units except sensor units are assigned an objective mean value $m^*_i = 0$. During learning, the agent applies the rule in Equation~\ref{eq:learning} for adjusting its means and correlations to the assigned values.
Each simulation step, the units of the Ising model are updated in a sequential random order using Glauber dynamics (Equation  \ref{eq:Glauber}).

The first embodiment of the network consists of the Mountain Car environment \cite{moore_efficient_1990}. This environment is a classical testbed in reinforcement learning depicting an under-powered car that must drive up a steep hill (Figure \ref{fig:Embodiment}.B). Since gravity is stronger than the car's engine, the vehicle must learn to gain potential energy by driving to one hill before the car is able to make it to the goal at the top of the opposite hill (see Methods). The neural network receives the speed of the system as an input and controls the force of the car's engine as an output.
The second embodiment consists of a double pendulum or `Acrobot' \cite{sutton_generalization_1996}, which has to coordinate the movements of two connected links to lift its weight (Figure \ref{fig:Embodiment}.C, see Methods). The neural network receives the speed of the first pendulum and controls the torque applied on the joint between the two pendulums.

For sizes $N_h=1,2,4,8,16,32,64$, we train 10 agents in both environments, applying the learning rule from Equation~\ref{eq:learning}, with $\eta=0.01$.
Note that agents during learning have no other explicit goal other than adjusting the correlations of the system to a random sample extracted from the probability distribution in Figure \ref{fig:Correlations}.
In Supplementary Videos S1 and S2 we can observe an example of the behaviour of agents with $N=64$ after training.
In Figure~S2.C-D we can see the distribution of correlations of an agent with $N_h=64$ (the result is similar for all agents and sizes) and Figure~S2.E-F shows the error between the distribution in Figures~S2.C-D and \ref{fig:Correlations} for this agents. In Figures~S2.C-D we can observe how agents after learning display a correlation structure consistent with the correlation distribution $c(r) \propto 1/r^{\eta}$ of the lattice Ising model at criticality.

In the following section, we analyse whether the agents designed in this way are able to operate in their environment in a regime of criticality.

\section*{Results}
In this section, we analyse the behaviour of the neural controllers and the behavioural patterns of the agents with respect to the possibilities of their parameter space. 
The goal is to test if the learning rule proposed here is effective for driving the agent near critical points.
As we observe below, the 10 agents display quite similar behaviour for each environment, despite the fact that each one has learned different values of $c^*_{ij}$ and $J^*_{ij}$. We are interested in analysing if there is anything special about the configuration reached by the agents after learning. 

In order to compare the agents with other behavioural possibilities, we explore the parameter space by changing the parameter $\beta$ of the agents. Modifying the value of $\beta$ is equivalent to a global rescaling of the parameters of the agent transforming $h_{i} \leftarrow  \beta \cdot h_{i}$ and $J_{ij} \leftarrow \beta \cdot J_{ij}$, thus exploring the parameter space along one specific direction. That is, changing $\beta$ is just a way of testing one dimension in the parameter space of possible models.
First, we assess the presence of criticality in the neural controller of the agent. Specifically, we look for the presence of a continuous phase transition in which an order parameter of the system (the entropy) presents a sharp transition displaying a divergence of its derivative (the heat capacity).
Second, we analyze not only the behaviour of the neural controller but the agent as a whole, looking for behavioural transitions of the agents and divergences of the susceptibility of the agent's behaviour to parametrical changes.

\subsection*{Signatures of criticality in the neural controller}

In order to test whether the agents are being driven towards a critical point, we analyse signatures of critical behaviour in the neural controller of the agent. 
As we mentioned before, a sufficient indicator for criticality is the presence of a divergence of the heat capacity of the system (as in Figure~\ref{fig:Test-Ising}.A). A divergence in the heat capacity indicates the presence of a continuous phase transition presenting a critical point in which the system is maximally sensitive to parametrical changes. Unfortunately, when the neural controller is embodied in the Mountain Car and Acrobot environments, we can no longer access a formal description of the probability distribution of the agent-environment system, thus we cannot directly compute from the energy of the system values as the entropy or heat capacity of the system.
Nevertheless, we can still directly compute the entropy $H(x) = -\sum_x P(x) \log{P(x)}$ of any variable of the system by estimating its probability distribution $P(x)$ through simulations. In order to compute the entropy and the heat capacity of different variables, we simulate the agent's behaviour for $101$ values of $\beta$, logarithmically distributed in the interval $[10^{-1},10^1]$. We run the 10 agents for each embodiment during $10^6$ simulation steps, reseting the agent's position and state every $5\cdot 10^4$ simulation steps.

\begin{figure}
\begin{center}
 \includegraphics{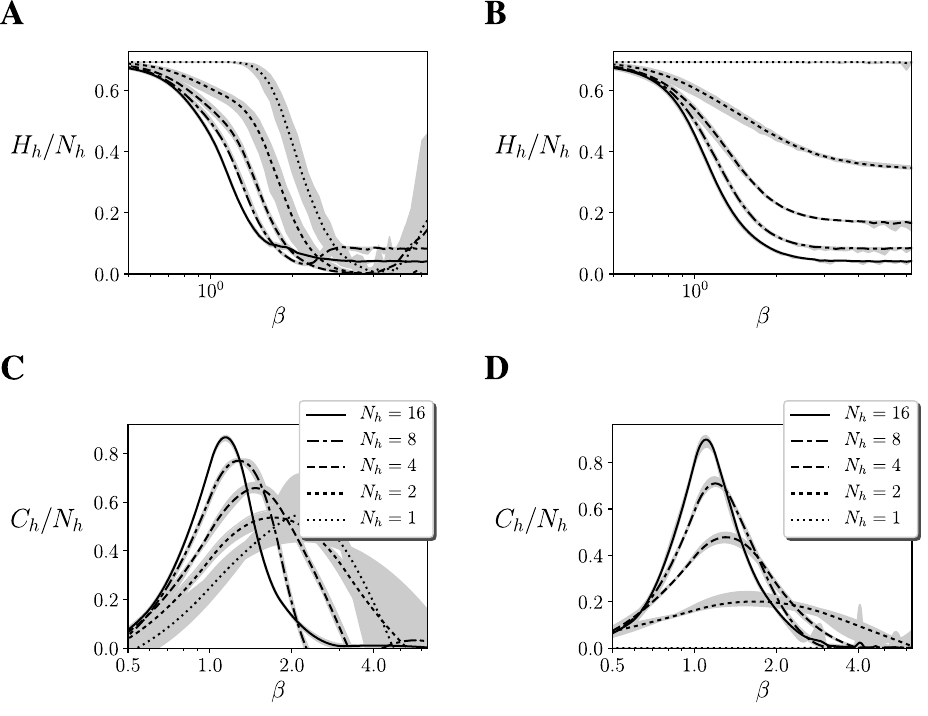}
\end{center}
\caption{\textbf{(A, B)} Entropy of the agent's neural controller for different sizes up to $N=22$ ($N_h=16$), for 10 different agents and 101 values of $\beta$. \textbf{(C, D)} Heat capacity of the agent's neurons, computed as $C(\beta) = - \beta \frac{\partial H(\beta)}{\partial \beta}$, where $H(\beta)$ is estimated using B-splines. Maximum and minimum values are shown by the grey area. The figures suggest that in both the Mountain Car (left) and Acrobot (right) embodiments the model presents a divergence of its heat capacity as the number of neurons $N$ increases, suggesting that the neural controller of the system is near a continuous phase transition.}
\label{fig:heat-capacity-n}
\end{figure}

First, we compute the entropy of the probability function of hidden neurons in the controller.  Due to computational constraints (the number of states of the probability distribution increases with $2^{N_h}$) we only compute the entropy in agents up to $N_h=16$ hidden units.
Displaying the entropy of the neurons for different values of $\beta$ we observe that the agents are near an order-disorder transition (Figure~\ref{fig:heat-capacity-n}.A,B). Larger sizes make the transition sharper and closer to the operating temperature $\beta=1$.
From the entropy of a variable, its heat capacity can be computed as $C(\beta) = -\beta \frac{\partial H(\beta)}{\partial \beta}$. 
From the computed $101$ values of entropy, $H(\beta)$ is estimated by fitting a curve using cubic B-splines \cite{dierckx_curve_1993} as indicated in the Methods section. 
In Figure~\ref{fig:heat-capacity-n}.C,D we observe how the system displays a similar divergence of the heat capacity of the neural network as the Ising model, suggesting that the robot's neural controller is at a critical point.

\begin{figure}
\begin{center}
 \includegraphics{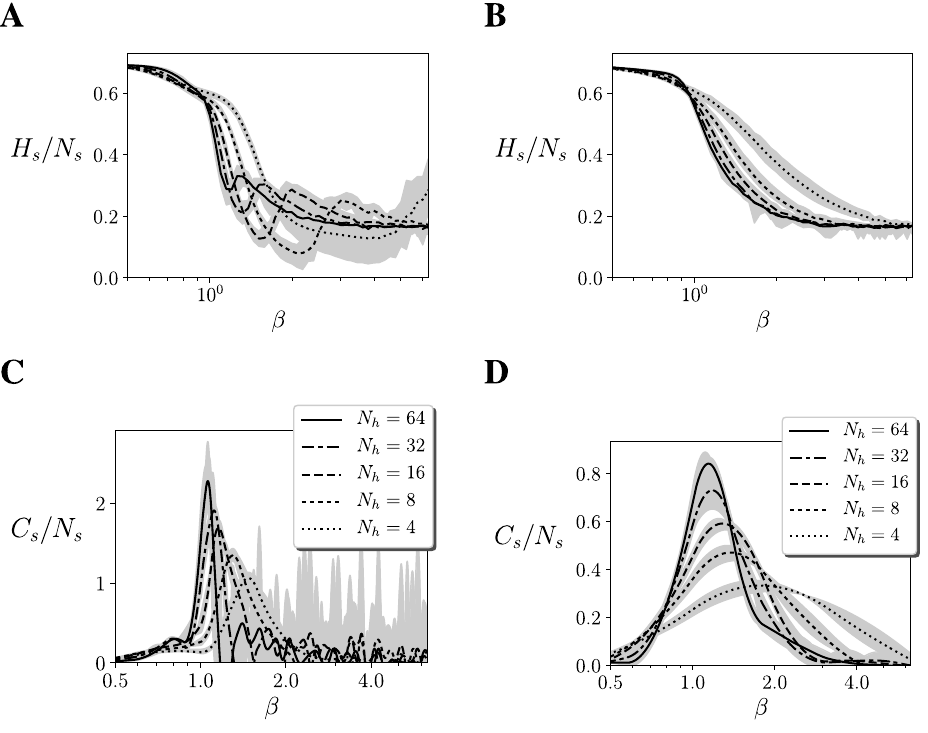}
\end{center}
\caption{\textbf{(A, B)} Entropy of the sensor units for different sizes up to $N=70$ ($N_h=64$), for 10 different agents and 101 values of $\beta$. \textbf{(C, D)} Heat capacity of the agent's neurons, computed as $C(\beta) = \beta \frac{\partial H(\beta)}{\partial \beta}$, where $H(\beta)$ is estimated using B-splines. Maximum and minimum values are shown by the grey area. The figures suggest that in both the Mountain Car (left) and Acrobot (right) embodiments the system presents a divergence of its heat capacity as the number of neurons $N$ increases, suggesting that the neural controller of the system is near a second order phase transition.}
\label{fig:heat-capacity-s}
\end{figure}

In order to confirm a divergence in the observed transition larger systems should be evaluated. Since evaluating the entropy of the hidden neurons is computationally infeasible for large sizes, we repeat the analysis for the state of the 4 sensor units of the network for a larger amount of hidden units.
In Figure~\ref{fig:heat-capacity-s} we show the entropy and heat capacity of the sensor units for sizes up to $N_h=64$ hidden neurons, where we can observe a similar picture than in Figure~\ref{fig:heat-capacity-n}, suggesting that the heat capacity diverges, and a second order transition takes place in the neural controller of the agent. 
These results suggest that the agent's neural controller is operating near a point of criticality, resembling a continuous phase transition, indicating that the neural controller self-organizes to present maximal sensitivity to changes in its parameter space.

\begin{figure}
\begin{center}
 \includegraphics{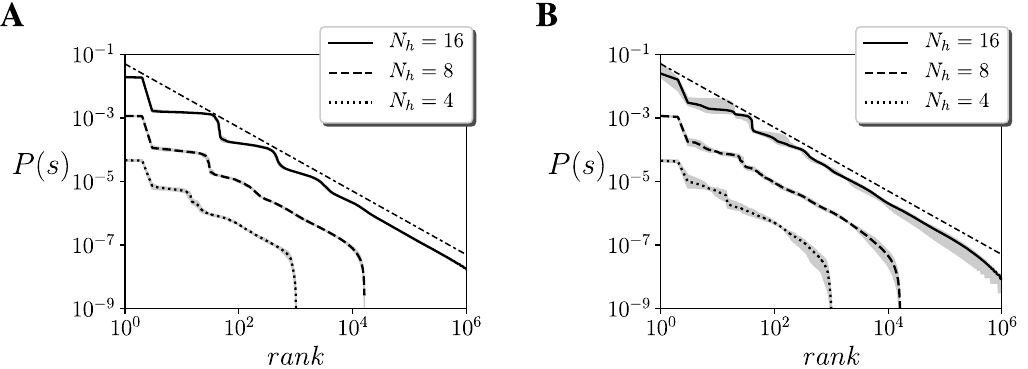}
\end{center}
\caption{Ranked probability distribution function of the neural network for 10 different agents and different sizes in \textbf{(A)} the Mountain Car and \textbf{(B)} the Acrobot embodiments. The real distribution is compared with a distribution following Zipf's law, (i.e. $P(s) = 1/rank$, dash-dotted line). We observe a good agreement between the model and Zipf's law, suggesting critical scaling.}
\label{fig:criticality}
\end{figure}

In addition to the presence of a continuous phase transition, a classical signature of criticality is the presence of power law distributions in the statistical descriptions of the states of a system. Unlike an isolated Ising model, our neural network is connected to an environment and  the probability distribution of the Ising neural controller is no longer described by Equation~\ref{eq:Ising}. Thus, we compute the probability distribution of the system by counting the occurrence of each state of the units $s$ to compute the probability distribution of the Ising model $P(s)$. We simulate the system at $\beta=1$ for $10^8$ simulation steps. As in training, the agent's position and state are reset every $5\cdot 10^4$ simulation steps.
We observe that all agents approximately follow Zipf's law for the Mountain Car (Figure~\ref{fig:criticality}.A) and Acrobot embodiments (Figure~\ref{fig:criticality}.B), with error bars in a very narrow range.
The power-law distribution of neural activation patterns suggests that the neural controller of the agents is operating near a critical point. We have to note that the sole occurrence of a power law is generally insufficient to assess the presence of criticality and may arise naturally in some non-equilibrium conditions. Nevertheless, together with the apparent divergence of the heat capacity it supports the idea that the neural controller of the agents might be poised near a critical state.

\subsection*{Behavioural transitions in the parameter space}
What does it imply for the agent to adapt to be poised near a critical point? It should be remarked here that agents are given no explicit goal. They only tend to adapt to behavioural patterns maintaining a distribution of correlations randomly sampled from the distribution shown in Figure~\ref{fig:Correlations}. To explore this issue, we examine the different behavioural modes of the agent while exploring its parameter space by changing the value of $\beta$. The behaviour of the Mountain Car can be described just by its horizontal position $x$ and speed $v$ at different moments of time. As well, the horizontal and vertical positions of the tip of the Acrobot's links is a good description of its behaviour.

\begin{figure}
\begin{center}
 \includegraphics{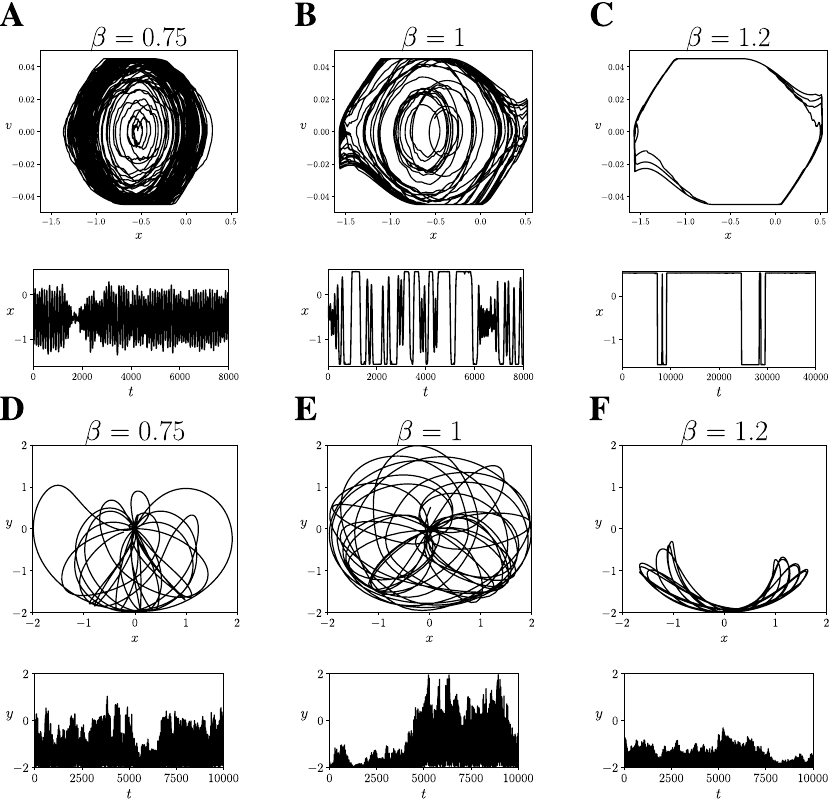}
\end{center}
\caption{Transition in the behavioural regime of the agents with $N_h=64$. We show the behaviour of two individual agents with different values of $\beta$ for the Mountain Car (\textbf{A}, \textbf{B}, \textbf{C}) and Acrobot (\textbf{D}, \textbf{E}, \textbf{F}) embodiments. We observe that $\beta=1$ is a transition point between two modes of behaviour in both agents.}
\label{fig:transitions}
\end{figure}

In Figure~\ref{fig:transitions}.A-C we can observe the behaviour of a Mountain Car agent with $N_h=64$ for $\beta=\{0.75,1,1.2\}$, respectively. We observe that for values of $\beta$ lower than the operating temperature, the agents are not able to reach the top of the mountain. On the other hand, when $\beta$ is higher, the agents present more `rigid' trajectories going from one  mountain peak to the other. At $\beta=1$ the agent is able to reach the top of the mountain (note that the peaks of the mountain are located at $x=-\pi/2$ and $x=\pi/6$) while displaying larger behavioural diversity.
Similarly, in Figure~\ref{fig:transitions}.D-F, we observe that the Acrobot agent with $N_h=64$ at $\beta=1$ displays a diverse range of behaviours, being able to reach the top of the plane while, when $\beta$ is lowered or increased, it drifts to other behavioural modes in less diverse regimes. Although only one agent is represented for each environment, the results of Figure~\ref{fig:transitions} are similar in all agents and sizes.

\begin{figure}
\begin{center}
 \includegraphics{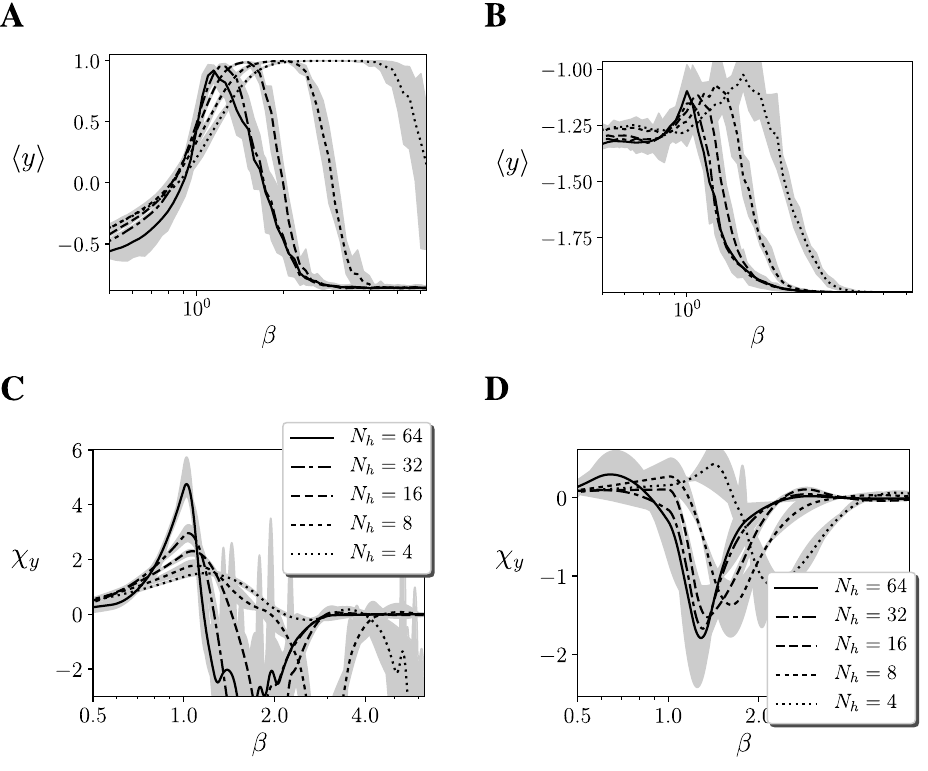}
\end{center}
\caption{\textbf{(A, B)} Mean height $\langle y \rangle$ of the agents for different sizes up to $N=70$ ($N_h=64$), for 10 different agents and 101 values of $\beta$. \textbf{(C, D)} Susceptibility the agent's behaviour, computed as $\chi_y(\beta) = \beta \frac{\partial \langle y \rangle}{\partial \beta}$, where $\langle y \rangle$ is estimated using B-splines. Maximum and minimum values are shown by the grey area. The figures suggest that in both the Mountain Car (left) and Acrobot (right) embodiments the behaviour of the agent presents a sharp transition around $\beta=1$.}
\label{fig:susceptibility}
\end{figure}

To get a more general picture of different agents and sizes, we can analyse the behavioural transitions in relevant variables of the agent environment systems.
Furthermore, we analyse whether the behaviour of the agent, and not only its neural controller, is near a critical point.
For inspecting this, we calculate the mean height of agents $\langle y \rangle$ in our simulations (Figure~\ref{fig:susceptibility}.A,B), and compute the susceptibility of this height value as  $\chi_y(\beta) = \beta \frac{\partial \langle y \rangle}{\partial \beta}$ (Figure~\ref{fig:susceptibility}.C,D). 
In this case, the susceptibility appears to increase monotonically with size when size is doubled, even if these increases are not as uniform as in Figures~\ref{fig:heat-capacity-n}  and \ref{fig:heat-capacity-s}. Although further tests of criticality could validate if criticality is also found in the whole agent-environment system, the figures suggest that the continuous phase transition of the neural controller corresponds with a sharp transition in the agent's behaviour.

\section*{Discussion}
Recapitulating the main ideas presented so far, we have tested how, by taking a set of correlations chosen at random from a distribution generated by a lattice Ising model at a critical point, we can construct a new model that appears to be also near a critical point in its parameter space. Moreover, if an embodied agent maintains these correlations using a simple learning rule while interacting with its environment -- as a sort of organizational homeostasis -- the agent neural controller seems to be driven to a critical point, which coincides with behavioural transitions in the agent-environment parameter space. 
Due to computational limitations for estimating the probability distribution of the embodied Ising networks, we only calculate the entropy and heat capacity of neural controllers up to $N_h=16$ hidden neurons (and $N=22$ total neurons) and approximate larger sizes computing the entropy and heat capacity of $N_s=4$ sensor neurons for neural controllers up to $N=70$ total neurons. These results are still far from the thermodynamic limit and further tests should confirm the results presented here for larger sizes. Nevertheless, in all cases we observe a clear diverging tendency of the heat capacity every time the size of the system is doubled.
Tests for larger sizes could be performed by designing environments that can be described by a Gibbs distribution, avoiding the combinatorial explosion in computing the heat capacity of the system by calculating it to be directly from the energy of the whole agent-environment system.

These results suggest the possibility that criticality could be diagnosed and induced directly from the maintenance of a given distribution of correlations rather than modelling a precise mechanistic structure. Also, our results show that criticality could be generated by quite simple mechanisms only relying on local information, maintaining specific correlations around a given value. Here we have implemented the mechanism as a simple Boltzmann Learning process, but other rules could have the same effect, such as the combination of Hebbian and anti-Hebbian tendencies in specific ratios or other simple mechanisms.
In our model, we only require the system to maintain a distribution of relations between the components of the system. 
This connects with systemic approaches to biology interested not in specific or intrinsic components of biological systems but in the networks of relations and processes \cite{bernard_introduction_1957, rosen_relational_1972, ashby_principles_1962}. It is also in line with notions of relational invariance in Piaget's approach to functional invariants in cognitive development \cite{piaget_biology_1974} or Maturana and Varela's ideas of autopoietic machines, defined as homeostatic systems that maintain  their own organization constant as a network of relations between components \cite{maturana_autopoiesis_1980}.

Assuming a similar systemic perspective, we have derived learning rule intended to drive a system towards critical points by maintaining an invariant structure of correlations roughly defined by a critical exponent $1/r^\eta$. 
Our approach assumes a different point of view on self-organized criticality, in which the distribution of correlations is not the consequence of criticality in a specific topology but the cause driving an indeterminate topology to what appears to be a critical point. 
The question now is whether imposing connections derived from a $1/r^\eta$ function is a strong assumption or implies particularly exigent circumstances. We do not think so, since power law functions can be naturally generated by simple rules of preferential attachment favouring `rich-get-richer' cumulative inequalities \cite{greenwood_inquiry_1920}, or directly as a natural consequence of certain geometries of space (e.g. gravitation laws\cite{barrow_new_2002}).

Our model only assumes that a system is going to adapt in order to preserve an internal network of relations. It emphasizes the maintenance of organizational structures capable of reproducing the behaviour of living systems without relying on internal models of the external source of sensory input. 
This contrasts with other approaches which have focused on understanding criticality as a strategy to effectively represent a complex and variable external world \cite{hidalgo_information-based_2014}, for example studying criticality in predictive coding or deep learning architectures dealing with complex inputs  \cite{friston_perception_2012, lin_critical_2016}. In those cases, an internalist view is assumed, where the neural controller represents structures of an external world, whose complexity may be the cause of critical activity in the neural controller. 
Instead, our approach is agnostic  in terms of the inputs or the external world of an organism, and deals only with how an agent rearranges its internal structures facing different environments. 

The agents presented here are not specifically designed for a particular problem. In simple terms, our agents generate (preserving the same internal neural organization) a wide variability and richness of behaviours (avoiding both disorder and explosive and indiscriminate propagation) that permits them to explore the space of parameters and eventually achieve solutions that they were not designed to find. The empirical evidence of experiments shown here supports this idea. 
A parallel could be established with the concept of play, which can be understood as a `rule-breaker' activity of the constraints of a stable and self-equilibrating regime of behaviours which has no concrete goals \cite{di_paolo_horizons_2010}. 
A model as the one presented here could be used for exploring life-like autonomous behaviour without the need for explicit internal representations, goals, or rule-based behaviour. 
Conceptual models of critical activity based on the maintenance of a system's relational invariants could help in the development of  a synthetic route towards the exploration of adaptive and embodied criticality.

% \noindent Please note: Abbreviations should be introduced at the first mention in the main text – no abbreviations lists. Suggested structure of main text (not enforced) is provided below.

% \section*{Introduction}
% 
% The Introduction section, of referenced text\cite{Figueredo:2009dg} expands on the background of the work (some overlap with the Abstract is acceptable). The introduction should not include subheadings.
% 
% \section*{Results}
% 
% Up to three levels of \textbf{subheading} are permitted. Subheadings should not be numbered.
% 
% \subsection*{Subsection}
% 
% Example text under a subsection. Bulleted lists may be used where appropriate, e.g.
% 
% \begin{itemize}
% \item First item
% \item Second item
% \end{itemize}
% 
% \subsubsection*{Third-level section}
%  
% Topical subheadings are allowed.
% 
% \section*{Discussion}
% 
% The Discussion should be succinct and must not contain subheadings.

\section*{Methods}

% \footnotesize
% \subsection{Environments}

\paragraph{Mountain Car.} This environment consists in a car with mass $m$ moving along a one-dimensional environment. In this environment, the agent moves its position in an horizontal axis $x$, limited to an interval of $[-1.5\pi/3,0.5\pi/3]$. Each horizontal position represents a point in an environment with two mountains, whose height is defined as $y=0.55+0.45\sin(3x)$. The velocity in the horizontal axis is updated each time step as $v(t+1)=v(t) + 0.001 a - 0.0025 \cos(3x)$, where $a$ is the action of the motor which can be either ${-1,0,1}$, impulsing the car with a force $F=m a$.
The inputs $I_i$ fed to the sensor units are defined as an array of 4 units, which are assigned the instantaneous velocity of the car discretized into an array of 4 bits. Each input $I_i$ is assigned a value of $1$ if its corresponding bit is active and $-1$ otherwise.

\paragraph{Acrobot.} The Acrobot is a two-link planar robot composed of two pendulums joined at their tip, with a motor applying a torque in clockwise or counterclockwise directions on the joint between the two links. The position of the system is defined by the angles of both pendulums $\theta_1$ and $\theta_2$, whose behaviour is defined by the following system of differential equations:
\begin{equation}
\begin{split}
\ddot \theta_{1} = -(d_{2} \ddot  \theta_{2} + \phi_{1}) / d_{1} \\
\ddot \theta_2=(m_2l_{c2}^2 + I_2 - \frac{d_2^2}{d_1})^{-1} (\tau + \frac{d_2}{d_1}\phi_1 - \phi_2)\\
d_{1} = m_{1} l_{c1}^2 + m_{2} (l_{1}^2 + l_{c2}^2 +\\+ 2 l_{1} l_{c2} \cos(\theta_{2})) + I_{1} + I_{2} \\
d_{2} = m_{2} (l_{c2}^2 + l_{1} l_{c2} \cos(\theta_{2})) + I_{2} \\
\phi_{2} = m_{2} l_{c2} g \cos(\theta_{1} + \theta_{2} - \pi / 2) \\
\phi_{1} = - m_{2} l_{1} l_{c2} \dot \theta_{2}^2 \sin(\theta_{2}) -\\- 2 m_{2} l_{1} l_{c2} \dot \theta_{2} \dot \theta_{1} \sin(\theta_{2}) +\\
+ (m_{1} l_{c1} + m_{2} l_{1}) g \cos(\theta_{1} - \pi / 2) + \phi_{2}
\end{split}
\end{equation}
where $\tau$ is the torque applied to the system which can be either ${-1,0,1}$, $m_1=m_2=m$ is the mass of the links, $l_1=l_2=1$ is the length of the links and $l_{c1}=l_{c2}=0.5$ are the lengths to the center of mass of the links, $I_1=I_2=1$ are the moments of inertia of the links and $g=9.8$ is the gravity. As well, variables $d_1$ and $d_2$ are the total moments of inertia of each link, and $\phi_1$ and $\phi_2$ are linked to the potential energy of the system

Similarly to the Mountain Car, the inputs fed to the sensor units in the Acrobot embodiment are defined as an array defined with 4 sensor units, encoding the angular speed of the first link with binary values (encoding active and inactive bits as +1 and -1). 

\paragraph{Task difficulty.} In order to make the tasks challenging, we set the maximum velocity allowed to the Mountain car to $\pm 0.045$ (typically is set to $0.07$) and the mass of the Acrobot links to $m=1.75$ (typically $m=1$). These parameters are designed to make it difficult for agents controlled by neural networks with random parameters solve the task (reaching the top of the environment), having success rates of  $6.1\%$  for the Mountain Car and $3.1\%$ for the 
Acrobot. Success rates were evaluated by simulating 1000 neural controllers with random parameters (sampled from a uniform distribution in the range $[-2,2]$). The Mountain Car was simulated for 1000 simulation steps starting from a random position in the valley between $[0.4,0.6]$, and was considered successful when reached the maximum position at least once. The Acrobot was simulated for 5000 simulation steps from the bottom position (angles and angular speeds between $[-0.1,01]$) and was consider successful if reached a vertical position higher than $1.8$.

\paragraph{Training.} 
During training, agents are initialized in the starting random positions in the bottom of their environments ($x\in [0.4,0.6]$ and $v=0$ for the Mountain Car and $\theta_1,\theta_2,\dot\theta_1,\dot\theta_2  \in [-0.1,01]$ for the Acrobot). 
The state of the neural network is randomized and the initial parameters $h_i$ and $J_{ij}$ are set to zero. 
Agents are simulated for $1000$ trials of $5000$ steps, computing each trial the values of $m_{i}^m$ and $c_{ij}^m$ and applying Equation~\ref{eq:learning} at the end of the trial. The agent's position and state are reset every $5\cdot 10^4$ simulation steps

\paragraph{Code availability.}
The source code implementing the learning rule in the different examples is freely available at \url{https://github.com/MiguelAguilera/Adaptation-to-criticality-through-organizational-invariance}.

\paragraph{Curve interpolation.} The heat capacity of hidden and sensor units was computed by interpolating entropy curves using cubic B-splines \cite{dierckx_curve_1993} using \emph{scipy} function \emph{splrep} with as smoothing coeficient of 1. Different coeficients were tested with similar results.

\bibliography{CriticalLearningII}

% \noindent LaTeX formats citations and references automatically using the bibliography records in your .bib file, which you can edit via the project menu. Use the cite command for an inline citation, e.g.  \cite{Figueredo:2009dg}.

\section*{Acknowledgements}

Research was supported in part by the Spanish National Programme for Fostering Excellence in Scientific and Technical Research project PSI2014-62092-EXP, by project TIN2016-80347-R funded by the Spanish Ministry of Economy and Competitiveness and the \mbox{UPV/EHU} post-doctoral training program \mbox{ESPDOC17/17}.

\section*{Author contributions statement}

% Must include all authors, identified by initials, for example:
M.A. conceived and conducted the experiments. M.A. and M.B. analysed the results and wrote the manuscript. 

\section*{Competing interests}

The authors declare no competing interests.

% \section*{Additional information}
% 
% To include, in this order: \textbf{Accession codes} (where applicable); \textbf{Competing financial interests} (mandatory statement). 
% 
% The corresponding author is responsible for submitting a \href{http://www.nature.com/srep/policies/index.html#competing}{competing financial interests statement} on behalf of all authors of the paper. This statement must be included in the submitted article file.

% \begin{figure}[ht]
% \centering
% \includegraphics[width=\linewidth]{stream}
% \caption{Legend (350 words max). Example legend text.}
% \label{fig:stream}
% \end{figure}
% 
% \begin{table}[ht]
% \centering
% \begin{tabular}{|l|l|l|}
% \hline
% Condition & n & p \\
% \hline
% A & 5 & 0.1 \\
% \hline
% B & 10 & 0.01 \\
% \hline
% \end{tabular}
% \caption{\label{tab:example}Legend (350 words max). Example legend text.}
% \end{table}

% Figures and tables can be referenced in LaTeX using the ref command, e.g. Figure~\ref{fig:stream} and Table \ref{tab:example}.

\end{document}